\documentclass[twocolumn,showpacs,preprintnumbers,superscriptaddress,amsmath,amssymb]{revtex4}
\usepackage{amsfonts,amssymb,amscd,amsmath}
\usepackage{graphicx}

 \newcommand\la{\langle}
 \newcommand\ra{\rangle}
 \newcommand\beq{\begin{equation}}
 
 \newcommand\eeq{\end{equation}}
 \newcommand\beqn{\begin{eqnarray}}
 \newcommand\eeqn{\end{eqnarray}}

\def\im{{\rm Im}}

\def\fm{\,\mbox{fm}}
\def\GeV{\,\mbox{GeV}}

\def\Pom{{\bf I\!P}}

\def\lsim{\mathrel{\rlap{\lower4pt\hbox{\hskip1pt$\sim$}}
    \raise1pt\hbox{$<$}}}         
\def\gsim{\mathrel{\rlap{\lower4pt\hbox{\hskip1pt$\sim$}}
    \raise1pt\hbox{$>$}}}         

\begin{document}
\title{Azimuthal anisotropy of direct photons}
 \author{B.Z.~Kopeliovich}
 \affiliation{Departamento de F\'\i sica y Centro de Estudios Subat\'omicos,
Universidad T\'ecnica Federico Santa Mar\'\i a, Casilla 110-V, Valpara\'iso,
Chile}
 \affiliation{Joint Institute for Nuclear Research, Dubna, Russia}
 \author{H.J. Pirner}
 \affiliation{Institute for Theoretical Physics, University of Heidelberg,
Philosophenweg 19, D-69120 Heidelberg, Germany}
 \author{A.H. Rezaeian$^1$}
 \author{Iv\'an Schmidt$^1$}

\date{\today}
 \begin{abstract}
 The electromagnetic bremsstrahlung produced by a quark interacting with
nucleons or nuclei is azimuthally asymmetric. In the light-cone dipole
approach this effect is related to the orientation dependent dipole cross
section. Such a radiation anisotropy is expected to contribute to the
azimuthal asymmetry of direct photons in $pA$ and $AA$ collisions, as well as
in DIS and in the production of dileptons.
  \end{abstract}
\maketitle
\date{\today}

\section{Introduction}

Direct photons, i.e. photons not from hadronic decay, are of
particular interest, since they do not participate in the strong
interaction and therefore carry undisturbed information about the
dynamics of the primary hard collision.

Here we present the basic color-dipole formalism for calculating the
azimuthal distribution of direct photons radiated by a quark interacting
either with a nucleon or nuclear targets. For this purpose we further
develop the dipole approach proposed in \cite{hir,kst1} for
electromagnetic bremsstrahlung by a quark interacting with nucleons and
nuclei. This technique can be applied to dilepton \cite{joerg,krtj,baier}
and prompt photon \cite{jamal,kprs} production in $pp$, $pA$ and heavy ion
collisions. It can be also used for calculating the azimuthal angle
dependence in the radiation of dileptons, or in deep inelastic scattering.

An azimuthal asymmetry appears due to dependence of the interaction
of a dipole on its orientation. Indeed, a colorless $\bar qq$ dipole
is able to interact only due to the difference between the impact
parameters of $q$ and $\bar q$ relative to the scattering center. If
$\vec b$ is the impact parameter of the center of gravity of the
dipole, and $\vec r$ is the transverse separation of the $q$ and
$\bar q$, then the dipole interaction should vanish if $\vec r
\perp\vec b$, but should be maximal if $\vec r
\parallel\vec b$. One can see this on a simple example
of dipole interacting with a quark in Born approximation. The
partial elastic amplitude reads,
 \beqn
&& \im f^q_{\bar qq}(\vec b,\vec r) = \frac{2}{9\pi^2} \int
\frac{d^2q\,d^2q'\,\alpha_s(q^2)\alpha_s(q'^2)} {(q^2+\mu^2)(q'^{2}+\mu^2)}\, 
\nonumber\\ &\times&
\left[e^{i\vec q\cdot(\vec b+\vec r/2)}- e^{i\vec q\cdot(\vec b-\vec
r/2)}\right]
\left[e^{i\vec q^{\,\prime}\cdot(\vec b+\vec r/2)}- e^{i\vec
q^{\,\prime}\cdot(\vec b-\vec r/2)}\right],
\nonumber\\
\label{2}
 \eeqn
 Here we assume for the sake of simplicity that $q$ and $\bar q$ have equal
longitudinal momenta, i.e. they are equally distant from the dipole center
of gravity. The general case of unequal sharing of the dipole momentum is
considered later in (\ref{300}). We introduced in (\ref{2})  an effective
gluon mass $\mu$ which takes into account confinement and other possible
nonperturbative effects. 

Integrating in (\ref{2}) with a fixed $\alpha_s$ we arrive at,
 \beq
\im f^q_{\bar qq}(\vec b,\vec r) = \frac{8\alpha_s^2}{9}
\left[K_0\left(\mu\left|\vec b+\frac{\vec r}{2}\right|\right) -
K_0\left(\mu\left|\vec b-\frac{\vec r}{2}\right|\right)\right]^2,
 \label{4}
 \eeq
 where $K_0(x)$ is the modified Bessel function. This
expression explicitly exposes a correlation between $\vec r$ and
$\vec b$: the two terms cancel each other if $\vec b\cdot\vec r=0$.

\section{Direct photons: dipole representation}

The radiation of direct photons, which in the parton model looks
like a Compton process $gq\to\gamma q$, in the target rest frame
should be treated as electromagnetic bremsstrahlung by a quark
interacting with the target. In the light-cone dipole approach the
transverse momentum distribution of photon bremsstrahlung by a quark
propagating interacting with a target $t$ (nucleon, $t=N$, or nucleus,
$t=A$) at impact parameter $\vec b$, can be written in the factorized
form \cite{kst1},
 \begin{eqnarray}
&&\frac{d \sigma^{qT\to\gamma X}(b,p,\alpha)}
{d(ln \alpha)\,d^{2}{p}\,d^{2}{b}}=\frac{1}{(2\pi)^{2}}
\sum_{in,f}\int d^{2}{r}_{1}d^{2}{r}_{2}
e^{i \vec{p}\cdot
(\vec{r}_{1}-\vec{r}_{2})}\nonumber\\
&\times&\phi^{\star}_{\gamma q}(\alpha, \vec{r}_{1})
\phi_{\gamma q}(\alpha, \vec{r}_{2})
F_t(\vec b,\alpha\vec{r}_{1},\alpha\vec{r}_{2},x). \label{m1}
 \end{eqnarray}
 Here $\vec p$ and $\alpha=p_\gamma^+/p_q^+$ are the transverse and fractional
light-cone momenta of the radiated photon, $\phi_{\gamma
q}(\alpha,\vec{r})$ is the light-cone distribution amplitude for the
$q\gamma$ Fock component with transverse separation $\vec r$, and
$F_t(\vec b,\alpha\vec{r}_{1},\alpha\vec{r}_{2},x)$ is an effective
partial amplitude to be discussed below. The product of the
distribution amplitudes, summed in (\ref{m1}) over initial and final
polarizations of the quark and photon, reads \cite{kst1},
 \begin{eqnarray}
&&\sum_{in,f}\phi^{\star}_{\gamma q}
(\alpha, \vec{r}_{1})\phi_{\gamma q}(\alpha, \vec{r}_{2})
= \frac{\alpha_{em}}{2\pi^{2}}m^2_{q}\alpha^{2}
\nonumber\\ &\times&
\Bigl\{\alpha^2 K_{0}(\alpha m_q r_{1})
K_{0}(\alpha m_q r_{2})\nonumber\\
&+&[1+(1-\alpha)^{2}]
\frac{\vec{r}_{1}.\vec{r}_{2}}{r_{1}r_{2}}
K_{1}(\alpha m_q r_{1})K_{1}(\alpha m_q r_{2})
\Bigr\}.
\label{wave}
 \end{eqnarray}
 Here $m_q$ is the effective quark mass, which is in fact an infra-red cutoff
parameter, and can be adjusted to photoproduction data \cite{kst2},
or shadowing \cite{krt1}, and whose value is $m_q\approx 0.2\GeV$.

In equation (\ref{m1}) the effective partial amplitude $F_t(\vec
b,\alpha\vec{r}_{1},\alpha\vec{r}_{2},x)$ is a linear combination of $\bar qq$
dipole partial amplitudes at impact parameter $b$,
 \begin{eqnarray}
F_t(\vec b,\alpha\vec{r}_{1},\alpha\vec{r}_{2},x)&=&
\im\Bigl[
f^t_{q\bar{q}}(\vec b,\alpha \vec r_{1},x)+
f^t_{q\bar{q}}(\vec b,\alpha \vec r_{2},x)
\nonumber\\
&-&
f^t_{q\bar{q}}(\vec b,\alpha(\vec{r}_{1}-
\vec{r}_{2}), x)\Bigr]\,,
\label{sig}
\end{eqnarray}
 where $x$ is Bjorken variable of the target gluons.

\section{Azimuthal asymmetry in quark-nucleon collisions}

In the case of a nucleon target ($t=N$), the partial elastic
amplitude $f^N_{\bar qq}(\vec b,\vec r)$ of interaction of the $\bar
qq$ dipole with a proton at impact parameter $\vec b$, is related to
the dipole cross section as,
 \beq
\sigma^N_{\bar qq}(r)=2\int d^2b\,
\im f^N_{\bar qq}(\vec b,\vec r).
\label{60}
 \eeq
 where $\sigma^N_{q\bar{q}}(r)$ is the total cross section of a $\bar qq$ - proton
collision. Here and further on, unless specified otherwise, the dipole cross
section and partial amplitudes implicitly depend on the Bjorken variable $x$ of the
target gluons.

The cross section $\sigma^N_{\bar qq}(r)$ has been rather
well determined by data on deep-inelastic scattering \cite{gbw}.
With this input, and using Eq.~(\ref{m1}), one can calculate the
inclusive differential cross section of direct photon emission. This
was done in \cite{kprs} for $pp$ collisions, with results in good
agreement with data.

Using the partial elastic amplitude $f^N_{\bar qq}(\vec b,\vec r)$
one can also calculate the differential elastic cross section of
dipole-nucleon scattering. Neglecting the real part, the amplitude
reads,
 \beqn
\frac{d\sigma^{(\bar qq)N}_{el}(r)}{dq_T^2} &=&
\frac{1}{4\pi}
\left|\int d^2b\,e^{i\vec q_T\cdot\vec b}
\im f^N_{\bar qq}(\vec b,\vec r)\right|^2
\nonumber\\ &\approx&
\frac{[\sigma^N_{\bar qq}(r)]^2}{16\pi}\,
\exp\left[-B^{(\bar qq)N}_{el}(r)q_T^2\right].
\label{70}
 \eeqn
 In the second line of this equation we rely on the small-$q_T$ approximation.
This defines the forward slope of the differential cross section,
which can be calculated as,
 \beq
B^{(\bar qq)N}_{el}(r)={1\over2}
\left\la s^2\right\ra =
\frac{1}{\sigma^N_{\bar qq}(r)}
\int d^2s\,s^2\,\im f^N_{\bar qq}(\vec s,\vec r).
\label{75}
 \eeq
 The slope for small-dipole-proton elastic scattering was measured in diffractive
electroproduction of $\rho$-mesons at high $Q^2$ at HERA \cite{diff-rho}.
The measured slope, $B^{(\bar qq)N}_{el}(r)\approx 5\GeV^{-2}$, agrees with the
expected value $B^{(\bar qq)N}_{el}(r)\approx B^{pp}_{el}/2$.

The objective of this paper is the azimuthal asymmetry of photon radiation. First of
all, we calculate the asymmetry of the cross section Eq.~(\ref{m1}) for quark-nucleon
collisions. The only vector available for such asymmetry is the impact parameter $\vec
b$, and therefore we should trace a correlation between the vectors $\vec p$ and $\vec
b$. The popular correlation function is defined as,
 \beqn
v^{qN}_2(b,p,\alpha)&=&\la\hat v_2\ra _{\phi_{p}}=
2\left\la \left(\frac{\vec p\cdot\vec b}
{p b}\right)^2\right\ra_{\phi_{p}} - 1
\nonumber\\ &=&
\frac{\int_0^{2\pi} d\phi_p\,\hat v_2\,
\frac{d \sigma^{qT\to\gamma X}(b,p,\alpha)}
{d(ln \alpha)\,d^{2}{p}\,d^{2}{b}}}
{\int_0^{2\pi} d\phi_p\,
\frac{d \sigma^{qT\to\gamma X}(b,p,\alpha)}
{d(ln \alpha)\,d^{2}{p}\,d^{2}{b}}}\,,
\label{80}
 \eeqn
 where the averaging is performed integrating in (\ref{m1}) over the azimuthal
angle $\phi_p$ of the transverse momentum $\vec p$.

\section{Radiation produced by a quark propagating through a nucleus}

In this case the partial amplitude to be used in (\ref{sig}), for a $\bar qq$
dipole colliding with a nucleus at impact parameter $b$, reads,
 \beqn
\im f^A_{q\bar{q}}(\vec b,\vec r) &=&
1-\left[1-{1\over2A}\sigma^N_{q\bar{q}}(r)
\tilde T_A(\vec b,\vec r)\right]^A
\nonumber\\ &\approx&
1-\exp\left[-{1\over2}\sigma^N_{q\bar{q}}(r)
\tilde T_A(\vec b,\vec r)\right].
\label{20}
 \eeqn
 The effective nuclear thickness $\tilde T_A$ is defined as
\cite{mine},
 \beq
\tilde T_A(\vec b,\vec r)=\frac{2}{\sigma^N_{\bar qq}(r)}\int d^2s\,
\im f^N_{\bar qq}(\vec s,\vec r)\,
T_A(\vec b+\vec s)\,,
\label{40}
 \eeq
 where the nuclear thickness function is defined as an 
integral of the nuclear density along the particle trajectory, 
$T_A(b)=\int_{-\infty}^\infty dz\rho_A(b,z)$.

Calculating $v_2^{qA}(b,p,\alpha)$,  we can average over $\phi_p$,
 \beq
\left\la \left(\frac{\vec p\cdot\vec b}
{p b}\right)^2\right\ra_{\phi_{p}}\propto
\int\limits_0^{2\pi} d\phi_p\,
\left(\frac{\vec p\cdot\vec b}
{p b}\right)^2\,
\frac{d \sigma^{qA\to\gamma X}(b,p,\alpha)}
{d(ln \alpha)\,d^{2}{p}\,d^{2}{b}}\,,
\label{80a}
 \eeq
 analytically. Instead of integration over direction of $\vec p$ at fixed
$\vec b$, one can integrate over direction of $\vec b$ at fixed $\vec p$.  
The advantage of such a replacement is obvious: all the $b$-dependence in
(\ref{m1}) is located in the effective amplitude $F^A$ and it has an explicit
and simple form.

Indeed, the mean value of $s^2$ is according to (\ref{75}) $\la s^2\ra\approx
0.4\fm^2$, which is much smaller than the heavy nucleus radius squared,
$R_A^2$. Therefore we can expand $T_A(\vec b+\vec s)$ as
 \beq
T_A(\vec b+\vec s)=
T_A(b) +\frac{\vec s\cdot\vec b}{b}T_A^\prime(b)
+{1\over2}\left(\frac{\vec s\cdot\vec b}{b}\right)^2
T_A^{\prime\prime}(b)+...
\label{100}
 \eeq
 Correspondingly, the partial amplitude (\ref{20}) can be expanded as,
 \beqn
&&\im f^A_{q\bar{q}}(\vec b,\vec r) \approx
1-\exp\left[-{1\over2}\sigma^N_{q\bar{q}}(r)T_A(b)\right]
\nonumber\\ &\times&
\Biggl\{1-\frac{1}{b}\,T_A^\prime(b)\,
\gamma_1(\vec b,\vec r)
\nonumber\\ &-&
\frac{1}{2b^2}\,\left[
T_A^{\prime\prime}(b)\gamma_2(\vec b,\vec r)-
T_A^{\prime^2}(b)\,\gamma_1^2(\vec b,\vec r)
\right]\Biggr\},
\label{120}
 \eeqn
 where
 \beq
\gamma_n(\vec b,\vec r)=
\int d^2s\,\im f^N_{q\bar{q}}(\vec s,\vec r)(\vec s\cdot\vec b)^n.
\label{140}
 \eeq

Integrating the amplitude (\ref{120}) together with $\hat v_2$ over $\phi_b$
we find that the first two terms in the curly brackets in (\ref{120}) give
zero, and the rest is,
 \beqn
&&\im\tilde f^A_{q\bar{q}}(b,\vec r) \equiv
\frac{1}{2\pi}\int\limits_0^{2\pi} d\phi_b\,
\im f^A_{q\bar{q}}(\vec b,\vec r)\,
\hat v_2(\phi_b)
\nonumber\\ &=&
e^{-{1\over2}\sigma^N_{q\bar{q}}(r)T_A(b)}
{1\over4}
\Bigl[T_A^{\prime\prime}(b)\,g(r)
-T_A^{\prime^2}(b)\,h(r)\Bigr].
\label{160}
 \eeqn
 Here
 \beq
g(r)=\int d^2s\,
\im f^N_{\bar qq}(\vec s,\vec r)\,
\left[2\,\frac{(\vec p\cdot\vec s)^2}{p^2}-s^2\right];
\label{180}
 \eeq
 \beqn
h(r)&=&\int d^2s_1 d^2s_2\,
\im f^N_{\bar qq}(\vec s_1,\vec r)
\im f^N_{\bar qq}(\vec s_2,\vec r)
\nonumber\\ &\times&
\left[2\,\frac{(\vec s_1\cdot\vec p)
(\vec s_2\cdot\vec p)}{p^2}-
(\vec s_1\cdot\vec s_2)\right].
\label{200}
 \eeqn

 Eq.~(\ref{160}) shows that the azimuthal asymmetry is strongly enhanced at the
nuclear periphery. Indeed, at small impact parameters the amplitude
Eq.~(\ref{160}) is suppressed by the factor
$\exp[-{1\over2}\sigma^N_{q\bar{q}}(r)T_A(b)]$, and moreover,
$T_A^\prime(b)\approx -2\rho_0\,b/\sqrt{R_A^2-b^2}$ and
$T_A^{\prime\prime}(b)\approx 2\rho_0\,R_A^2/(R_A^2-b^2)^{3/2}$ are
vanishingly small and peak at the periphery ($\rho_0\approx 0.16\fm^{-3}$ is
the central nuclear density). The smallness of both the amplitude and azimuthal
asymmetry justifies also the expansion made in (\ref{120}).

As far as the partial amplitudes $f^N_{\bar qq}(\vec s,\vec r)$ and their
asymmetric part Eq.~(\ref{160}) are known, one can calculate the azimuthal
asymmetry of photons radiated in quark-nucleus collisions,
 \beqn
&& v_2^{qA}(b,p,\alpha)\,\frac{d \sigma^{qA\to\gamma X}(b,p,\alpha)}
{d(ln \alpha)\,d^{2}{p}\,d^{2}{b}}=
\sum_{in,f}\int d^{2}{r}_{1}d^{2}{r}_{2}
\nonumber\\&\times&
e^{i \vec{p}\cdot
(\vec{r}_{1}-\vec{r}_{2})}
\phi^{\star}_{\gamma q}(\alpha, \vec{r}_{1})
\phi_{\gamma q}(\alpha, \vec{r}_{2})
\tilde F^A(b,\alpha\vec{r}_{1},\alpha\vec{r}_{2}).
\label{220}
 \eeqn
 where
 \beqn
\tilde F^A(b,\alpha\vec{r}_{1},\alpha\vec{r}_{2})&=&
\im \tilde f^A_{q\bar{q}}(b,\alpha r_{1})+
\im \tilde f^A_{q\bar{q}}(b,\alpha r_{2})
\nonumber\\
&-&
\im \tilde f^A_{q\bar{q}}\bigl(b,\alpha|\vec{r}_{1}-
\vec{r}_{2}|\bigr).
\label{240}
 \eeqn
 Notice that the cross section in the left hand side of Eq.~(\ref{220}) can also
be calculated without using the expansion (\ref{100}), relying on the eikonal
approximation, $\tilde T_A(\vec b,\vec r)\approx T_A(b)$, which is known to
be quite accurate for heavy nuclei.

\section{Partial dipole amplitude $f^N_{\bar qq}(\vec b,\vec 
r)$}

The next step is to model the partial dipole amplitude $f^N_{\bar qq}(\vec b,\vec
r)$. An azimuthal asymmetry can only emerge if the amplitude
$f^N_{\bar qq}(\vec b,\vec r)$ contains a correlation between the vectors $\vec b$
and $\vec r$. If such a correlation is lacking, the functions Eqs.~(\ref{180})
and (\ref{200})  are equal to zero. A model for $f^N_{\bar qq}(\vec b,\vec r)$ 
having no $\vec b-\vec r$ correlation was proposed in \cite{kt}.

It is rather straightforward to calculate the partial amplitude within the two gluon
exchange model \cite{klz},
 \beqn
\im f^N_{\bar qq}(\vec b,\vec r) &=&
\frac{2}{3\pi^2}
\int \frac{d^2q\,d^2q'\,\alpha_s(q^2)\alpha_s(q'^2)}
{(q^2+\mu^2)(q'^{\,2}+\mu^2)}
\nonumber\\ &\times&
e^{i\vec b\cdot(\vec q-\vec q^{\,\prime})}\,
\Bigl(1 -
e^{i\vec q\cdot\vec r}\Bigr)\Bigl(1 -
e^{-i\vec q^{\,\prime}\cdot\vec r}\Bigr)
\nonumber\\ &\times&
\left[F_N(\vec q-\vec q^{\,\prime})-
F_N^{(2q)}(\vec q,\vec q^{\,\prime})\right],
\label{260}
 \eeqn
 where $F_N(k)=\la\Psi_N|\exp(i\vec k\cdot\vec\rho_1)|\Psi_N\ra$ is the
nucleon form factor, and $F_N^{(2q)}(\vec q,\vec
q^{\,\prime})=\la\Psi_N|\exp[i\vec q\cdot\vec\rho_1-i\vec
q^{\,\prime}\cdot\vec\rho_2]|\Psi_N\ra$ is the so called two-quark nucleon
form factor. Both can be calculated using the three valence quark nucleon wave
function $\Psi_N(\vec\rho_1,\vec\rho_2,\vec\rho_3)$.

An effective gluon mass $\mu$ is introduced in (\ref{260}) in order to
imitate confinement. We fix its value at $\mu=m_\pi$ in order to reproduce the large
hadronic cross sections.

The Born amplitude is unrealistic since leads to an energy
independent dipole cross section $\sigma_{\bar qq}(r,x)$. This dipole cross section
has been well probed by
measurements of the proton structure function at small Bjorken $x$ at HERA,
and was found to rise towards small $x$, with an $x$ dependent steepness.
In fact, it can be expressed via the unintegrated gluon density
${\cal F}(x,q^2)$,
 \beq
\sigma(r,x)=\frac{4\pi}{3}\int \frac{d^2q}{q^4}\,
\left(1-e^{-i\vec q\cdot\vec r}\right)\,
\alpha_s(q^2)\,{\cal F}(x,q^2)\,.
\label{280}
 \eeq

 Analogously, the partial amplitude for dipole-nucleon elastic 
scattering at impact parameter $\vec b$ between the centers of 
gravity of the dipole and nucleon reads,
 \beqn
&&\im f^N_{\bar qq}(\vec b,\vec r,\beta)=\frac{1}{12\pi}
\int\frac{d^2q\,d^2q'}{q^2\,q'^2}\,\alpha_s\,
{\cal F}(x,\vec q,\vec q^{\,\prime})
e^{i\vec b\cdot(\vec q-\vec q^{\,\prime})}
\nonumber\\ &\times&
\left(e^{-i\vec q\cdot\vec r\beta}-
e^{i\vec q\cdot\vec r(1-\beta)}\right)\,
\left(e^{i\vec q'\cdot\vec r\beta}-
e^{-i\vec q'\cdot\vec r(1-\beta)}\right)\,
\,.
\label{300}
 \eeqn
 Here the dipole has transverse separation $\vec r$, fractional light-cone
momenta of the quark and antiquark, $1-\beta$ and $\beta$ respectively.
Since the radiated photon takes away fraction $\alpha$ of the quark
momentum, the corresponding dipole has $\beta=1/(2-\alpha)$. The 
impact parameter $\vec b$ of the dipole is the transverse distance 
from the target to the dipole center of gravity, which is shifted towards 
the fastest $q$ or $\bar q$ in accordance with (\ref{300}).

In (\ref{300}) $\alpha_s=\sqrt{\alpha_s(q^2)\alpha_s(q'^2)}$, and we
introduced the off-diagonal unintegrated gluon density ${\cal F}(x,\vec
q,\vec q^{\,\prime})$, which in the Born approximation limit takes the
form,
 \beqn
{\cal F}(x,\vec q,\vec q^{\,\prime}) &\Rightarrow&
{\cal F}_{Born}(\vec q,\vec q^{\,\prime})
\nonumber\\ &=&
\frac{4\alpha_s}{\pi}
\left[F_N(\vec q-\vec q^{\,\prime})-
F_N^{(2q)}(\vec q,\vec q^{\,\prime})\right].
\label{310}
 \eeqn

Besides, the partial elastic amplitude Eq.~(\ref{300}), should satisfy the
conditions Eqs.~(\ref{60}) and (\ref{75}).  For the dipole cross section we
rely on the popular saturated shape \cite{gbw} fitted to HERA data for
$F_2^p(x,Q^2)$ and we choose the following form of ${\cal F}(x,\vec q,\vec 
q^{\,\prime})$,
 \beqn
 {\cal F}(x,\vec q,\vec q^{\,\prime}) &=&
\frac{3\,\sigma_0}{16\,\pi^2\,\alpha_s}\ q^2\,q'^2\,R_0^2(x)
\nonumber\\ &\times&
{\rm exp}\Bigl[-{1\over8}\,R_0^2(x)\,(q^2+q'^2)\Bigr]
\nonumber\\ &\times&
{\rm exp}\bigl[-R_N^2(\vec q-\vec q^{\,\prime})^2/2\bigr]
\,,
 \label{320}
 \eeqn
 where $\sigma_{0}=23.03$ mb, $R_{0}(x)=0.4 \fm \times (x/x_{0})^{0.144}$
with $x_{0}=3.04\times 10^{-4}$ \cite{gbw} and $x=p/\sqrt{s}$ \cite{kprs}.
We assume here that the Pomeron-proton form factor has the Gaussian form,
$F^p_\Pom(k_T^2)=\exp(-k_T^2 R_N^2/2)$, so the slope of the $pp$ elastic
differential cross section is
$B^{pp}_{el}=2R_N^2+2\alpha^\prime_\Pom\ln(s/s_0)$, where
$\alpha^\prime_\Pom\approx 0.25\GeV^{-2}$ is the slope of the Pomeron
trajectory, $s_0=1\GeV^2$. $R_N^2\approx \la r_{ch}^2\ra/3$ is the part of
the slope of elastic cross section related to the Pomeron-proton form
factor.

With this unintegrated gluon density the partial amplitude Eq.~(\ref{300})
can be calculated explicitly,
\begin{widetext}
 \beq
\im f^N_{\bar qq}(\vec b,\vec r,x,\beta) =
\frac{\sigma_0}{8\pi B_{el}}\,
\Biggl\{\exp\left[-\frac{[\vec b+\vec r(1-\beta)]^2}{2B_{el}}\right] +
\exp\left[-\frac{(\vec b-\vec r\beta)^2}{2B_{el}}\right]
-2\exp\Biggl[-\frac{r^2}{R_0^2}-
\frac{[\vec b+(1/2-\beta)\vec r]^2}{2B_{el}}\Biggr]
\Biggr\},
\label{340}
 \eeq
\end{widetext}
 where $B_{el}(x)=R_N^2+R_0^2(x)/8$. This amplitude satisfies the
conditions Eqs.~(\ref{60}) and (\ref{75}). This expression also goes 
beyond the usual assumption that the dipole cross section is independent 
of the light-cone momentum sharing $\beta$. The partial amplitude 
Eq.~(\ref{340}) does depend on $\beta$, but this dependence disappears 
after integration over impact parameter $\vec b$.

\section{Numerical results}

\begin{figure}[thb] \centerline{\includegraphics[width=7 cm] 
{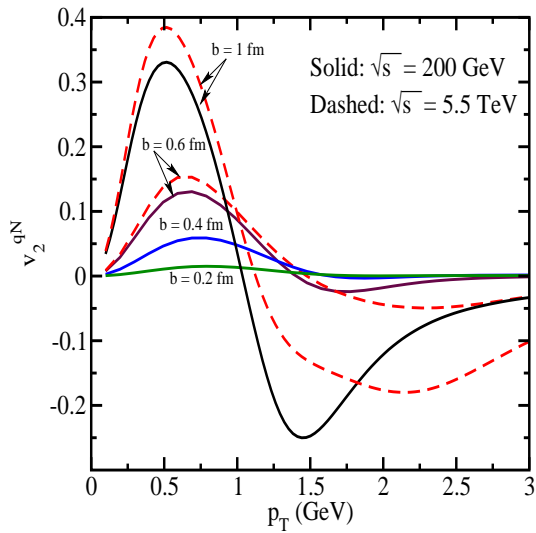}}
\caption{The anisotropy parameter $v_2^{qN}(b,p,\alpha)$ as function of $p$
calculated at $\alpha=1$ for different impact parameters $b$ and energies:
$\sqrt{s}=200\GeV$ (solid, $b=0.2,\ 0.4,\ 0.6,\ 1\fm$), and
$\sqrt{s}=5500\GeV$ (dashed, $b=0.6,\ 1\fm$).
 \label{fig-1}}
 \end{figure}

Now we are in a position to calculate $v_2^{qN}(b,p,\alpha)$. Examples of
quark-nucleon collisions radiating a photon, with $\alpha=1$ and at different impact
parameters and energies, are depicted in Fig.~\ref{fig-1}.
 The results show that the anisotropy of the dipole interaction rises with
impact parameter, reaching rather large values. As function of the transverse
momentum of the radiated photons, $v_2^{qN}(b,p,\alpha)$ vanishes at large
$p_T$. Such a behavior could be anticipated, since the interaction of
vanishingly small dipoles responsible for large $p$ is not sensitive to the
dipole orientation.

\begin{figure}[thb]
\centerline{\includegraphics[width=7 cm] {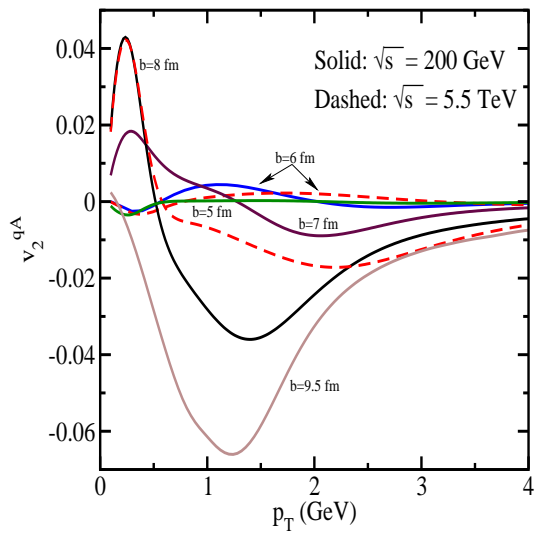}}
\caption{
 Azimuthal anisotropy of direct photons with $\alpha=1$ from quark-lead
collisions at different impact parameters as is labeled in the plot. Solid and
dashed curves correspond to the energies of RHIC ($b=5,\ 6,\ 7,\ 8,\ 
9.5\fm$), and LHC ($b=6,\ 8\fm$) respectively.
 \label{fig-2}}
 \end{figure}

The next step is calculating the azimuthal asymmetry in quark-nucleus collisions.
The results are plotted in Fig.~\ref{fig-2} as function of transverse momentum, at
different impact parameters and at the energies of RHIC and LHC.

 The first observation is the smallness of $v^{qA}_2$, which is suppressed an
order of magnitude compared to $v^{qN}_2$. At first glance this might look
strange, since the quark interacts with nucleons anyway. However, a quark
propagating through a nucleus interacts with different nucleons located at
different azimuthal angles relative to the quark trajectory. Their
contributions to $v_2^{qA}$ tend to cancel each other, restoring the azimuthal
symmetry. Such cancellation would be exact if the nuclear profile function
$T_A(b)$ were constant. We have a nonzero, but small $v_2^{qA}$ only due to
the variation of $T_A$ with $b$, i.e. the presence of finite first and second
derivatives, as was derived in Eq.~(\ref{160}).

The results of a numerical integration (without expansion (\ref{120})),
depicted in Figs.~\ref{fig-1}-\ref{fig-2}, also confirm the anticipation based
on Eq.~(\ref{160}) that the azimuthal asymmetry is enhanced on the nuclear
periphery.

\begin{figure}[t]
\centerline{\includegraphics[width=7 cm] {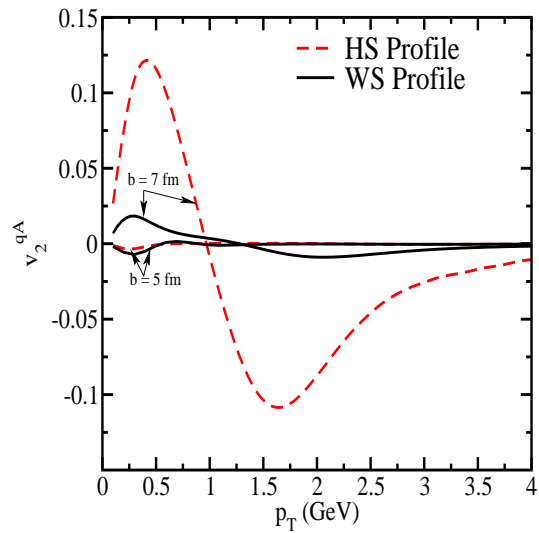}}
 \caption{Azimuthal anisotropy of direct photons with $\alpha=1$ from
quark-lead collisions at $b=5$ and $7\fm$. Solid and dashed curves are
calculated with Woods-Saxon (WS) and hard sphere (HS)  parametrizations of
nuclear density respectively.
 \label{fig-3}}
 \end{figure}

 We used the Woods-Saxon parametrization for nuclear density \cite{jager}.
The anisotropy of electromagnetic radiation appears only on the nuclear
periphery and according to (\ref{160}) is extremely sensitive to the behavior
of the nuclear thickness function at the very edge of the nucleus. Electron
scattering data, which is the main source of information about the electric charge
distribution in nuclei, is not sensitive to the neutron distribution, which is
known to be enlarged on the periphery. Therefore the details of the shape of
the density distribution on the nuclear surface are poorly known. As a simple
estimate of the theoretical uncertainty related to this problem one can use an
alternative parametrization of the nuclear density, such as  the simple
and popular hard sphere form, $\rho(r)=\rho_0\Theta(R_A-r)$. We compare in
Fig.~\ref{fig-3} the anisotropy parameters $v^{qA}_2(p,b,\alpha)$ calculated
with hard sphere (dashed) and Woods-Saxon (solid) parametrizations.
 As one could expect, the hard sphere density leads to a quite larger
anisotropy, since the derivatives of the nuclear profile function are much
sharper.

\section{Summary}

Summarizing, we extended the dipole description of electromagnetic radiation
\cite{hir,kst1} in quark nucleon and nucleus collisions to calculation of the
azimuthal angle distribution. This problem involves more detailed features of
the dipole amplitude, namely its dependence on dipole size and impact
parameter, as well as on their correlation. We propose a simple model
generalizing the unintegrated gluon density fitted to HERA data for the proton
structure function to an off-diagonal gluon distribution. The latter satisfies
all the imposed boundary conditions.

The developed theoretical tools can be applied to the calculation of the azimuthal
asymmetry in DIS and in Drell-Yan reactions on a proton, as well as
to the production of direct photons and Drell-Yan pairs in proton-nucleus and
heavy ion collisions.

\section*{Acknowledgments}
 This work was supported in part by Fondecyt (Chile) grants 1070517 and 1050589 and
by DFG (Germany) grant PI182/3-1. AHR is grateful to the hospitality of Hans
Pirner´s group at Heidelberg University where this work was started, and
acknowledges the financial support from the Alexander von Humboldt 
foundation.

\end{document}